\documentclass[showkeys,showpacs,superscriptaddress,prd]{revtex4}
\usepackage{amsmath}
\usepackage{amssymb}
\usepackage{graphicx}
\usepackage{float}

\begin{document}

\title{
Model of a spin-$1/2$ electric charge in $F\left( B^2\right)$ modified Weyl gravity
}

\author{
Vladimir Dzhunushaliev
}
\email{v.dzhunushaliev@gmail.com}
\affiliation{
Department of Theoretical and Nuclear Physics,  Al-Farabi Kazakh National University, Almaty 050040, Kazakhstan
}
\affiliation{
Institute of Experimental and Theoretical Physics,  Al-Farabi Kazakh National University, Almaty 050040, Kazakhstan
}
\affiliation{
National Nanotechnology Laboratory of Open Type,  Al-Farabi Kazakh National University, Almaty 050040, Kazakhstan
}
\affiliation{
Academician J.~Jeenbaev Institute of Physics of the NAS of the Kyrgyz Republic, 265 a, Chui Street, Bishkek 720071, Kyrgyzstan
}

\author{Vladimir Folomeev}
\email{vfolomeev@mail.ru}
\affiliation{
Institute of Experimental and Theoretical Physics,  Al-Farabi Kazakh National University, Almaty 050040, Kazakhstan
}
\affiliation{
National Nanotechnology Laboratory of Open Type,  Al-Farabi Kazakh National University, Almaty 050040, Kazakhstan
}
\affiliation{
Academician J.~Jeenbaev Institute of Physics of the NAS of the Kyrgyz Republic, 265 a, Chui Street, Bishkek 720071, Kyrgyzstan
}
\affiliation{
International Laboratory for Theoretical Cosmology, Tomsk State University of Control Systems and Radioelectronics (TUSUR),
Tomsk 634050, Russia
}

\begin{abstract}
Within $F\left( B^2\right)$ modified Weyl gravity, we consider a model of a spin-$1/2$ electric charge consisting of interior and exterior regions.
The interior region is determined by quantum gravitational effects
whose approximate description is carried out using Weyl gravity nonminimally coupled to a massless Dirac spinor field. The interior region is embedded in exterior Minkowski spacetime, and the joining surface is a two-dimensional torus. It is shown that mass, electric charge, and spin of the object suggested may be the same as those for a real electron.
\end{abstract}



\date{\today}

\maketitle

\section{Introduction}

One of the unresolved problems in fundamental physics is the problem of the electron's internal structure.
In classical physics, this problem manifests itself in the fact that a point charge possesses an infinite mass related to the energy
of the electrostatic field created by the charge. In quantum electrodynamics, a point electron leads to the occurrence of divergences in loop graphs.

Apparently, a first attempt to resolve this problem has been made by Abraham and Lorentz within the theory of an electron suggested in Refs.~\cite{ablo,ablo1}.
Another attempt to resolve the problem is Wheeler's idea of `charge without charge'~\cite{wheelbook} where an electric charge is regarded as a
wormhole containing force lines of an electric field. Within such model, the region where the force lines enter the wormhole looks like a negative charge,
and the region where the force lines emerge from the wormhole mimics a positive charge. Also, numerous attempts have been made to find regular solutions in classical and nonlinear electrodynamics both in the presence  and in the absence of the gravitational
field~\cite{Born:1934gh, heis, Markov:1970hv, Markov:1972sc, Bronnikov:1979ex, Datzeff:1980tg, Bonnor:1989iw,   Herr:1994, Zaslavskii:2010yi,  Finst:1999, Bronn:2000, Habib:2009, Zaslav:2010, Dzhunushaliev:2016bnh}.
One should also definitely mention the popular approach to the classical model of an electron using the Einstein-Dirac~\cite{Finster:1998ws}
and  Kerr-Newman~\cite{KNold,KNold1,KNold2,KNold3,KNold4,KNold5,KNold6,KNold7} solutions (an extensive bibliography on the subject can be found in Ref.~\cite{Burinskii:2011id}).
In Ref.~\cite{Davidson:1992kr}, invoking world-manifold gauge field theory, Dirac's idea of finite size electron is considered.
The idea that an electric monopole with internal magnetic monopole like-structure can be interpreted as a genuine electric monopole by an outside observer
was considered in Ref.~\cite{Davidson:1990ez}.

In the present paper we consider a model of an electron on the assumption that its internal structure is determined by quantum gravitational effects.
We suppose that such effects can be approximately described by modified Weyl gravity suggested in Ref.~\cite{Dzhunushaliev:2020dom}. 
We wish to note here that modified gravities are usually assumed to have the general relativistic limit, satisfy solar system tests, and give correct Newton's law
(see, e.g., Refs.~\cite{Nojiri:2010wj,Nojiri:2017ncd}). 
Within modified Weyl gravity under consideration, all this is not required, since we work in opposite direction: we suppose that  
modified Weyl gravity is some approximation in quantum general relativity. For this reason, it is not necessary to perform the aforementioned limit passages.
In other words, not general relativity is a limiting case of modified gravity but modified Weyl gravity is some approximation for quantum general relativity. 
This enables us to represent the electron as a composite object 
whose interior region is described by modified Weyl gravity and the exterior region~-- by Minkowski spacetime. In turn, in both regions, there is a Dirac spinor field that ensures the presence of a spin of $1/2$ in the system. To describe the interior region, we will use the solution obtained in  Ref.~\cite{Dzhunushaliev:2020dom} which is embedded in an exterior flat spacetime. In addition, notice that, since Weyl gravity (as well as modified Weyl gravity considered here) is conformally invariant, metrics differing by a conformal factor are the same solution. Physically this can be interpreted  as the presence in the interior region of fluctuations of the metric conformal factor.

\section{Spin-$1/2$ particles in $F(B^2)$ modified Weyl gravity}

To begin with, let us briefly review the results obtained in Ref.~\cite{Dzhunushaliev:2020dom}. It was shown there that in $F(B^2)$ modified Weyl gravity containing a Dirac spinor field, there are regular solutions with finite energy and spin $1/2$. The physical treatment of such $F(B^2)$ modified Weyl gravity consists in the assumption that this theory can approximately describe quantum gravitational effects in general relativity for large magnitudes of the scalar curvature $R$.

The action for modified Weyl gravity nonminimally coupled to a spinor field can be written in the form (hereafter we work in natural units
$\hbar = c = 1$)
\begin{equation}
	\mathcal S = \int d^4 x \sqrt{-g} \left[
	- \alpha_g C_{\alpha \beta \gamma \delta}
	C^{\alpha \beta \gamma \delta} +
	F \left( B^2 \right) \mathcal L_\psi
		\right] ,
\label{Weyl_10}
\end{equation}
where $\alpha_g$ is a dimensionless constant,
$
C_{\alpha \beta \gamma \delta} = R_{\alpha \beta \gamma \delta} +
\frac{1}{2} \left(
R_{\alpha \delta} g_{\beta \gamma} -
R_{\alpha \gamma} g_{\beta \delta} +
R_{\beta \gamma} g_{\alpha \delta} -
R_{\gamma \delta} g_{\alpha \beta}\right) + 
\frac{1}{6} R \left( 
	g_{\alpha \gamma} g_{\beta \delta} - 
	g_{\alpha \delta} g_{\beta \gamma}
\right) 
$ is the Weyl tensor,
$B^2 = B_{\mu \nu} B^{\mu \nu}$ is the Bach scalar curvature invariant,
$
	B_{\mu \nu} =
	2 C^{\alpha \phantom{\mu \nu} \beta}_
	{\phantom{\alpha} \mu \nu \phantom{\beta} ; \alpha \beta} +
	C^{\alpha \phantom{\mu \nu} \beta}_
	{\phantom{\alpha} \mu \nu \phantom{\beta}} R_{\alpha \beta}
$ is the Bach tensor,
   $F \left( B^2 \right)$ is an arbitrary function,   $L_\psi $ is the Lagrangian density of the massless spinor field $\psi$,
$$
	\mathcal L_\psi = \frac{i }{2} \left(
			\bar \psi \gamma^\mu \psi_{; \mu} -
			\bar \psi_{; \mu} \gamma^\mu \psi
		\right),
$$
which contains the covariant derivative
$
\psi_{; \mu} \equiv \left[\partial_{ \mu} +1/8\, \omega_{a b \mu}\left(
\gamma^a  \gamma^b- \gamma^b  \gamma^a\right)\right]\psi
$,
where $\gamma^a$ are the Dirac matrices in flat space (below we use the spinor representation of the matrices). In turn, the Dirac matrices in curved space,
$\gamma^\mu = e_a^{\phantom{a} \mu} \gamma^a$, are obtained using
the tetrad
$ e_a^{\phantom{a} \mu}$, and $\omega_{a b \mu}$ is the spin connection
[for its definition, see Ref.~\cite{Lawrie2002}, formula (7.135)].

The action \eqref{Weyl_10} and hence the corresponding theory are invariant under the conformal transformations
$$
	g_{\mu \nu} \rightarrow f^2(x^\alpha) g_{\mu \nu}, \quad
	\psi \rightarrow f^{-3/2}(x^\alpha) \psi ,
$$
where the function $f(x^\alpha)$ is arbitrary.

The corresponding set of equations in $F \left( B^2 \right)$ modified Weyl gravity is
\begin{eqnarray}
	\alpha_g B_{\mu \nu} &=&
	\frac{1}{4 \alpha_g} F \left( B^2 \right) T_{\mu \nu} ,
\label{Weyl_90}\\
	i \gamma^\mu \psi_{;\mu} &=& 0 ,
\label{Weyl_100}\\
	T_{\mu \nu} &=& \frac{i}{4} \left[
		\bar\psi \gamma_{\mu} \psi_{; \nu} +
		\bar\psi \gamma_\nu \psi_{;\mu} 	-
		\bar\psi_{; \mu}\gamma_{\nu} \psi -
		\bar\psi_{; \nu} \gamma_\mu \psi
	\right] .
\label{Weyl_110}
\end{eqnarray}
The solution to this set of equations is sought in the form
\begin{eqnarray}
	ds^2 &=& f^2 \left(
		t, \chi, \theta, \varphi
	\right) \left\{
		dt^2 - \frac{r^2}{4}
		\left[
			\left( d \chi - \cos\theta d \varphi\right)^2
			+ d \theta^2 + \sin^2 \theta d \varphi^2
		\right]
	\right\} = f^2 \left(
	t, \chi, \theta, \varphi
	\right) \left( dt^2 - r^2 dS^2_3 \right) ,
\label{solution_10}\\
	\psi_{m,n} &=&
	f^{-3/2} e^{-i \Omega t} e^{i n \chi} e^{i m \varphi}
	\begin{pmatrix}
		\Theta_1(\theta)  \\
		\Theta_2(\theta)  \\
		\Theta_3(\theta)  \\
		\Theta_4(\theta)
	\end{pmatrix} ,
\label{1_10}
\end{eqnarray}
where $m, n$, and $\Omega$ are free parameters,
$f \left( t, \chi, \theta, \varphi \right)$ is an arbitrary function, and
$dS^2_3$ is the Hopf metric on the unit sphere (for a detailed description of the Hopf fibration, see, e.g., Ref.~\cite{Dzhunushaliev:2020dom});
$\chi, \theta, \varphi$ are Hopf coordinates on the three-dimensional space related to the Cartesian coordinates
 $x, y, z$ as follows:
\begin{equation}
	x = r \frac{
		\sin \left( \frac{\theta}{2} \right)
		\cos \left(\frac{\varphi + \chi}{2}\right)
	}
	{1 - \cos \left( \frac{\theta}{2} \right) \sin \left(\frac{\varphi - \chi}{2}\right) },
\quad
	y = r \frac{
		\sin \left( \frac{\theta}{2} \right)
		\sin \left(\frac{\varphi + \chi}{2}\right)
	}
	{1 - \cos \left( \frac{\theta}{2} \right) \sin \left(\frac{\varphi - \chi}{2}\right) },
	\quad
	z = r \frac{
		\cos \left( \frac{\theta}{2} \right)
		\cos \left(\frac{\varphi - \chi}{2}\right)
	}
	{1 - \cos \left( \frac{\theta}{2} \right) \sin \left(\frac{\varphi - \chi}{2}\right) } .
\label{A_120}
\end{equation}

\begin{figure}[H]
\begin{minipage}[ht]{.49\linewidth}
\centering
	\includegraphics[width=.8\linewidth]{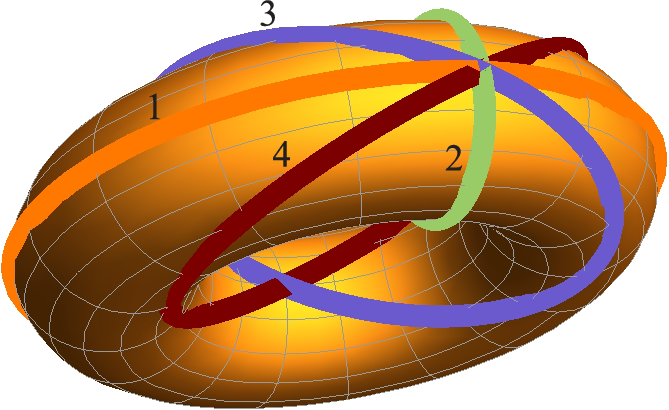}
\vspace{.7cm}
\caption{The circles on a torus:
1 and 2 are ordinary coordinate circles, 3 and 4 denote the Villarceau circles.
}
\label{torus_villar}
\end{minipage}
\begin{minipage}[ht]{0.49\linewidth}
\centering
	\includegraphics[width=1\linewidth]{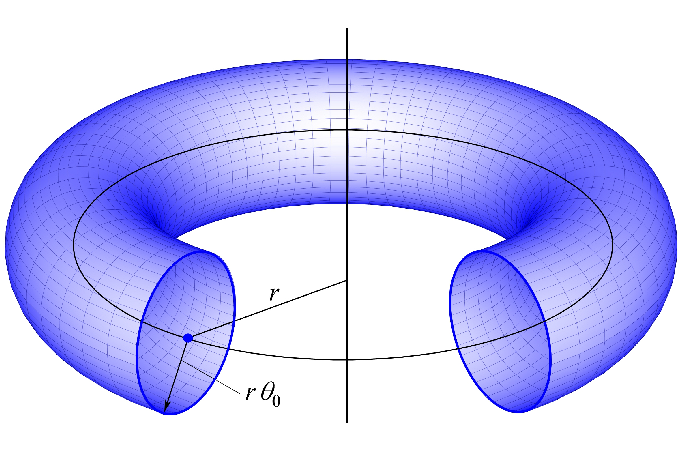}
\vspace{-1.1cm}
\caption{
The torus $T^2_{\text{join}}$ over which the interior solution with the metric~\eqref{solution_10} and with one of the spinors~\eqref{2_A_20}-\eqref{2_A_50}
or~\eqref{2_B_20} is embedded in exterior Minkowski spacetime.
}
\label{joining_torus}
\end{minipage}\hfill
\end{figure}

Usually, as coordinates on a torus, one chooses angular coordinates whose coordinate lines are circles on a torus depicted in Fig.~\ref{torus_villar}
as the circles 1 and 2. Hopf coordinates are the angles $\chi$ and $\varphi$ whose coordinate lines with
$\theta, \chi = \text{const}$ and $\theta, \varphi = \text{const}$ are shown by the Villarceau circles 3 and 4.

One can show that for the metric~\eqref{solution_10} the Bach tensor vanishes,
$
	B_{\mu \nu} = 0
$.
This enables us to simplify considerably the equations of $F \left( B^2 \right)$ modified Weyl gravity. Namely, by choosing
$$
	F \left( B^2 \right) = 0 ,
$$
the modified Bach equation~\eqref{Weyl_90} is identically satisfied and the only remaining equation is the Dirac equation~\eqref{Weyl_100}
(for details concerning such a choice of $F \left( B^2 \right)$ see Ref.~\cite{Dzhunushaliev:2020dom}).
This equation has two types of solutions.

\subsection{The case $m = n = 0$}

In this case we have the following solution
\begin{eqnarray}
	\Theta_{1, 2} &=&
	\frac{C}{\sqrt{\sin \theta} } , \quad
	\Theta_{2, 1} = \Theta_3 = \Theta_4 = 0 , \quad
	r \Omega = \frac{1}{2};
	\label{1_20}\\
	\Theta_{3, 4} &=& \frac{C}{\sqrt{\sin \theta} } , \quad
	\Theta_{4, 3} = \Theta_1 = \Theta_2 = 0 , \quad
	r \Omega = - \frac{1}{2},
\label{2_A_10}
\end{eqnarray}
where $C=\left(4 \pi r^3	\right)^{-1/2}$ is a normalization constant and the parameters $m$ and $n$ are taken to be zero.
Accordingly, the spinors~\eqref{1_10} can be written in the form
\begin{eqnarray}
	\psi_1^T &=& C
	\frac{e^{ - i t / (2 r)}}{f^{3/2}} \left\lbrace
		\frac{1}{\sqrt{\sin \theta} }, 0, 0, 0
	\right\rbrace ,
\label{2_A_20}\\
	\psi_2^T &=& C
		\frac{e^{- i t / (2 r)}}{f^{3/2}} \left\lbrace
		0, \frac{1}{\sqrt{\sin \theta} }, 0, 0
	\right\rbrace ,
\label{2_A_30}\\
	\psi_3^T &=& C \frac{e^{ i t / (2 r)}}{f^{3/2}} \left\lbrace
		0, 0, \frac{1}{\sqrt{\sin \theta} }, 0
	\right\rbrace ,
\label{2_A_40}\\
	\psi_4^T &=& C \frac{e^{ i t / (2 r)}}{f^{3/2}} \left\lbrace
		 0, 0, 0, \frac{1}{\sqrt{\sin \theta} }
	\right\rbrace ,
\label{2_A_50}
\end{eqnarray}
where the indices $1,2,3,4$ by $\psi$ correspond to four linearly independent solutions. These spinors satisfy the eigenvalue problem for the operator of the projection
of the total angular momentum on the $z$-axis:
\begin{equation}
	\left( n^i \hat{\tilde M}_i \right) \psi_{1, 3} =
	\frac{1}{2} \psi_{1, 3} ,
\quad
	\left( n^i \hat{\tilde M}_i \right) \psi_{2, 4} =
	- \frac{1}{2} \psi_{2, 4},
\label{2_A_60}
\end{equation}
where the operator of the projection of the total angular momentum on the unit vector
$n_i$ is $\hat{\tilde M}_i = \hat{\tilde L}_i + \hat{\tilde S}_i$ (its calculation and the description of the corresponding quantities are given in
Appendix~\ref{spin_oper}). Here, the tilde above the symbols denotes that the corresponding quantity is taken in Hopf coordinates.
Since we consider the projection on the $z$-axis, we have $n_i = (0, 0, 1)$.
Thus, the solutions~\eqref{2_A_20}-\eqref{2_A_50} describe the object possessing the projection of the spin on the $z$-axis
 equal to $\pm 1/2$.

\subsection{The case $m, n \neq 0$}

It was shown in Ref.~\cite{Dzhunushaliev:2020dom} that in this case the Dirac equations are split into two set of equations:
one for the unknown functions $\Theta_{1,2}$ and the other one -- for the functions $\Theta_{3,4}$. When
$\Theta_{1,2} \neq 0$ and $\Theta_{3,4} = 0$, the corresponding particular solutions can be represented in the form
\begin{equation}
	\left( \Theta_{1} \right)_{m,n} =
	\pm \left( \Theta_{2} \right)_{m,n} =
	C \sin^\alpha \left( \frac{\theta}{2} \right)
	\cos^\beta \left( \frac{\theta}{2} \right) , \quad
	\Theta_{3, 4} = 0 ,
\label{2_B_10}
\end{equation}
where
\begin{equation}	
\alpha = \pm\left(n+m\right) - \frac{1}{2}, \quad
	\beta = \pm \left(n-m\right) - \frac{1}{2} ,
\quad
	r \Omega_n = \frac{1}{2}\left(1\pm 4 n\right)
\label{Omega}
\end{equation}
and $C$ is a complex integration constant. These solutions form a discrete spectrum depending on two quantum numbers $m$ and $n$.

Accordingly, the solution~\eqref{1_10} can be written in the following form:
\begin{equation}
	\psi^T = C
	\frac{
		e^{-i \Omega t} e^{i n \chi} e^{i m \varphi}
	}{f^{3/2}}
	\sin^\alpha \left( \frac{\theta}{2} \right)
	\cos^\beta \left( \frac{\theta}{2} \right)
	\left\lbrace
		1, \pm 1, 0, 0
	\right\rbrace .
\label{2_B_20}
\end{equation}
Then the action of the operator of the projection of the total angular momentum $\left( n^i \hat{\tilde M}_i \right)$ on the $z$-axis on the spinor
\eqref{2_B_20} takes the form
\begin{equation}
	\left( n^i \hat{\tilde M}_i \right) \psi =
	C
	\frac{
		e^{-i \Omega t} e^{i n \chi} e^{i m \varphi}
	}{f^{3/2}}
	\sin^\alpha \left( \frac{\theta}{2} \right)
	\cos^\beta \left( \frac{\theta}{2} \right)
	\left\lbrace
		\frac{1}{2} + m + n, \pm \frac{1}{2} + m + n, 0, 0
	\right\rbrace .
\label{2_B_30}
\end{equation}
It is seen from this expression that the equality
$$
	\left( n^i \hat{\tilde M}_i \right) \psi = M_z \psi
$$
will be satisfied only if $m = - n$, where $M_z$ is an eigenvalue.

\section{Model of a spin-$1/2$ electric charge}
\label{eModel}

For brevity, we will henceforward call the spin-$1/2$ electric charge  considered in the present paper simply an `electron' (in quotation marks).
The `electron' consists of two regions: the exterior one, where the spacetime is flat, and the interior one,
where it is described by modified Weyl gravity with the metric~\eqref{solution_10}. Geometrically this metric describes a torus shown in Fig.~\ref{joining_torus}.
The conformal factor $f \left(	t, \chi, \theta, \varphi	\right)$ appearing in Eq.~\eqref{solution_10} is not determined
by the field equations;  it can be chosen arbitrarily. We will assume that inside the `electron' (i.e., inside the torus) all conformal factors are equiprobable,
and the metric in general fluctuates between all possible forms of $f \left(	t, \chi, \theta, \varphi	\right)$.
Below we consider one possible form of the conformal factor for which the spatial part of the metric~\eqref{solution_10}
describes a flat space.

\subsection{Metric \eqref{solution_10} with a flat spatial part
}
\label{flat}

To construct a model of a spin-$1/2$ electric charge, we note first that the spatial part of the metric
\begin{equation}
	ds^2 = \left[
	\frac{2}{
			2 - \sin \left( \frac{\theta + \varphi - \chi}{2} \right) +
			\sin \left( \frac{\theta - \varphi + \chi}{2} \right)
	}	
	\right]^2
	\left\{
		dt^2 - \frac{r^2}{4}
		\left[
			\left( d \chi - \cos\theta d \varphi\right)^2
			+ d \theta^2 + \sin^2 \theta d \varphi^2
		\right]
	\right\}
\label{2_10}
\end{equation}
is flat [see Eq.~\eqref{A_120}]:
$$
	dl^2 = dx^2 + dy^2 + dz^2 =
	\left[
		\frac{r}{
				2 - \sin \left( \frac{\theta + \varphi - \chi}{2} \right) +
				\sin \left( \frac{\theta - \varphi + \chi}{2} \right)
		}	
		\right]^2
		\left[
			\left( d \chi - \cos\theta d \varphi\right)^2
			+ d \theta^2 + \sin^2 \theta d \varphi^2
		\right] .
$$

The key idea of the model suggested here consists in that we cut out the interior region of the torus with $\pi - \theta_0 \leq \theta \leq \pi$ (where $\theta_0 \ll 1$)
from the spacetime with the metric~\eqref{2_10} and embed it in Minkowski spacetime over the surface $\theta = \pi - \theta_0$ (see Fig.~\ref{joining_torus}).
As a result, the spatial part of the metric inside and outside the torus is smoothly joined, since in both cases this is the same flat metric.
However, in this case the temporal component of the metric~\eqref{2_10} has a discontinuity,
\begin{equation}
	g_{tt} = \begin{cases}
		\left[
				\frac{2 r}{
					2 - \sin \left( \frac{\theta + \varphi - \chi}{2} \right) +
					\sin \left( \frac{\theta - \varphi + \chi}{2} \right)
				}
		\right]^ 2 & \text{ if } \pi - \theta_0 \leq \theta \leq \pi
	\\
		1  & \text{ if } \theta > \pi - \theta_0
	\end{cases}.
\label{2_30}
\end{equation}

Of course, for a real electron, there need in general be no such a discontinuity. Actually, there should be some transition region near the torus where the solutions are smoothly joined.
Inside the torus, quantum gravitational effects play an important role, and one can neglect them outside it.
In any case, the description of physical processes in the transition region requires the use of quantum gravity theory, which is absent at the moment.

\subsection{Metric \eqref{solution_10} with $f \left(	t, \chi, \theta, \varphi	\right) = 1$}

As an interior metric, one can also employ a metric whose spatial part is the Hopf metric
\begin{equation}
	ds^2 = dt^2 - \frac{r^2}{4}
		\left[
			\left( d \chi - \cos\theta d \varphi\right)^2
			+ d \theta^2 + \sin^2 \theta d \varphi^2
		\right] .
\label{3_b_10}
\end{equation}
In this case, when joining the interior and exterior solutions, the temporal component of the metric
 $g_{tt}$ does not already have a discontinuity on the torus $T^2_{\text{join}}$, but there is a discontinuity of the spatial components of the metrics,
\begin{equation}
	dl^2 = \begin{cases}
		\frac{r^2}{4}
			\left[
				\left( d \chi - \cos\theta d \varphi\right)^2
				+ d \theta^2 + \sin^2 \theta d \varphi^2
			\right] & \text{ if } \pi - \theta_0 \leq \theta \leq \pi
	\\
		f^2 \left(	t, \chi, \theta, \varphi	\right) \frac{r^2 }{4}
			\left[
				\left( d \chi - \cos\theta d \varphi\right)^2
				+ d \theta^2 + \sin^2 \theta d \varphi^2
		\right]  & \text{ if } \theta > \pi - \theta_0
	\end{cases},
\label{3_b_20}
\end{equation}
where now
$$
	f \left(	t, \chi, \theta, \varphi	\right) =
	\frac{2 }{
		2 - \sin \left( \frac{\theta + \varphi - \chi}{2} \right) +
		\sin \left( \frac{\theta - \varphi + \chi}{2} \right)
	} .
$$
Here, as in the case described in Sec.~\ref{flat}, in the place of the discontinuity of the metric on the torus $T^2_{\text{join}}$,
there must be a transition region whose description should be carried out using methods of quantum gravity.

Summarizing the results obtained in this section, the interior and exterior regions of the `electron' have been joined at the torus
 $T^2_{\text{join}}$ in two cases:
\begin{itemize}
\item When the spatial part of the metric~\eqref{2_30} inside the torus $T^2_{\text{join}}$ is flat, and the temporal component of the metric has a discontinuity in the region of joining.
\item When the spatial part of the metric inside the torus $T^2_{\text{join}}$ is the Hopf metric~\eqref{3_b_10}
and there is a discontinuity in the form~\eqref{3_b_20}. In this case, there is no discontinuity of the temporal component of the metric.
\end{itemize}

Thus, the `electron' under consideration is an object consisting of exterior and interior regions which are joined at the torus  $T^2_{\text{join}}$,
and the interior metric may in a certain sense be regarded  as a quantum superposition of states with the metrics~\eqref{2_10} and \eqref{3_b_10}.

\section{Mass, charge, and spin of the `electron'
}

Let us now calculate the total energy (mass) $W$, the charge $q$, and the projection of the total angular momentum $M_z$ on the $z$-axis of the `electron' under consideration:
\begin{align}
	W &= \int \limits_{V \subset T^2_{\text{join}}}\epsilon_\psi  d V\equiv\Omega
	\int \limits_{V \subset T^2_{\text{join}}} \psi^\dagger \psi d V \equiv
	\Omega \gamma ,
\label{4_10}\\
	q &= \int \limits_{V \subset T^2_{\text{join}}} j^0 dV =
	e_0 \int \limits_{V \subset T^2_{\text{join}}} \psi^\dagger \psi dV =
	e_0 \gamma ,
\label{4_20}\\
	M_z &= s_z
\label{4_30}
\end{align}
where $s_z = \pm 1/2$ is an eigenvalue of the operator $n^i \hat{\tilde M}_i$ from the eigenproblem \eqref{2_A_60}, and we have also used the expression for the current density
$j^\mu = e_0 \bar \psi \gamma^\mu \psi$,
and the integration over the interior region of the torus $T^2_{\text{join}}$ appearing in the above formulae is defined as
$$
	\int \limits_{V \subset T^2_{\text{join}}} \ldots  d V =
	\frac{r^3}{8} \int\limits_{\pi - \theta_0}^\pi
	\int\limits_{0}^{2 \pi} \int\limits_{0}^{2 \pi}  \ldots
	\sin \theta d \theta d \chi d \varphi .
$$
In calculating $W$, we use the following definition of the energy density of the spinor field:
$$
	\epsilon_\psi =i \bar \psi \gamma^0 \dot \psi - L_{\psi}.
$$
We emphasize that the integration in Eqs.~\eqref{4_10} and \eqref{4_20} is carried out only over the volume of the torus $T^2_{\text{join}}$.

\subsection{The case $m = n = 0$}
\label{m_n_0}

In this case
\begin{equation}
	\gamma = \int \limits_{V \subset T^2_{\text{join}}} \psi^\dagger \psi d V = \frac{\pi^2 r^3}{2} C^2 \theta_0 \ll 1.
\label{4_A_10}
\end{equation}
Here, we have taken into account that the torus $T^2_{\text{join}}$ is determined by the magnitude of the angle
$\theta = \theta_0 \ll 1$ (see Sec.~\ref{flat}). Then Eqs.~\eqref{4_10}-\eqref{4_30} yield
\begin{align}
	W &= \frac{\gamma}{2 r} ,
\label{4_A_20}\\
	q &= e_0 \gamma ,
\label{4_A_30}\\
	M_z & = \pm \frac{1}{2}.
\label{4_A_40}
\end{align}
In Eq.~\eqref{4_A_20}, we have used  $r \Omega = 1 /2$; in Eq.~\eqref{4_A_40}, we have employed the expression~\eqref{2_A_60},
with the plus sign taken for the indices  $(1, 3)$ and the minus sign~-- for the indices $(2, 4)$.

If one chooses $W = m_e$ (the electron mass) and $q = e$  (the electron charge), then the `electron' will have the electron mass and charge,
and the projection of its total angular momentum on the $z$-axis will be equal to $\pm 1/2$. To provide such characteristics of the `electron', it is necessary that
\begin{equation}
	\gamma = 2 m_e r = 2 \frac{r}{\lambda_c} \ll 1, \quad e_0 = \frac{e}{\gamma} = e \frac{\lambda_c}{2 r} \gg e .
\label{4_A_50}
\end{equation}
Here $\lambda_c$ is  the Compton wavelength of the electron and we suppose that the characteristic size of the `electron' must be much smaller than
 $\lambda_c$  [see Eq.~\eqref{4_A_50}].

\subsection{The case $m,n \neq 0$}
\label{m_n_neq_0}

In this case the quantity $\gamma$ is defined as
\begin{equation}
	\gamma = \int \limits_{V \subset T^2_{\text{join}}} \psi^\dagger \psi d V = \frac{ \pi^2 r^3}
	{2^{1 + 2 \beta} (\beta + 1)} C^2 \theta_0^{2 + 2 \beta} \ll 1 ,
	\quad \beta = \pm \left(
		n - m - \frac{1}{2}
	\right),
\label{4_B_10}
\end{equation}
where the signs $\pm$ correspond to the signs $\pm$ in the expression~\eqref{2_B_10}. Accordingly, Eqs.~\eqref{4_10}-\eqref{4_30} yield
\begin{align}
	W &= \Omega_n \gamma = \frac{1 \pm 4 n}{2 r} \gamma ,
\label{4_B_20}\\
	q &= e_0 \gamma ,
\label{4_B_30}\\
	M_z &= \pm \frac{1}{2} .
\label{4_B_40}
\end{align}
Consider the case with the plus sign in the expression \eqref{4_B_20}. For convenience, we introduce the constant
\begin{equation}
	C_{m,n}^2 = \frac{ \pi^2 r^3 C^2}	{2^{1 + 2 \beta} (\beta + 1)}.
\label{norm_const}
\end{equation}
Then the parameter
$$
	\gamma = C_{m,n}^2 \theta_0^{3 + 2 m - 2 n} .
$$
If we again take $W = m_e$ and $q = e$, then we get the following relations for the `electron':
\begin{align}
	2 m_e r &= \frac{2 r}{\lambda_c} =
	(1 + 4 n) C_{m,n}^2 \theta_0^{3} \ll 1 ,
\label{4_B_50}\\
	\frac{e}{e_0} &= \gamma = C_{m,n}^2 \theta_0^{3} \ll 1 .
\label{4_B_60}
\end{align}
The condition~\eqref{4_B_50} is necessary so that the size of the `electron' will be much smaller than the Compton wavelength of the electron $\lambda_c$.

Let us now consider in more detail the parameter $\gamma$, whose numerical value is determined from Eq.~\eqref{4_B_10}. The condition~\eqref{4_B_50} means that the size of the torus $T^2_{\text{join}}$ (i.e., the size of the interior region of the `electron') should be very small compared to $\lambda_c$. Then, taking into account that the radius of the axial circle of the torus $T^2_{\text{join}}$ is equal to $r$, the radius of the cross-section of the torus for small $\theta_0$ can be estimated as $\approx r \theta_0$ (see Fig.~\ref{joining_torus}). Hence, one can estimate the characteristic radius of the `electron' as $r$, and its cross-section radius as $r \theta_0$. If we assume that, say,  $\lambda_c \gg r \gg r \theta_0$, then we have $1 \gg \gamma \gg r \theta_0/\lambda_c$. Then, if the cross-section radius is chosen to be, say, $r \theta_0 \sim l_{Pl}$, this yields $1 \gg \gamma  \gg 10^{-23}$.

Finally, consider the relationships between $W = m_e$, $q = e$ on the one hand and $r, e_0$ on the other [see Eqs.~\eqref{4_B_50} and~\eqref{4_B_60}]. In our case
$m_e \ll m_0 \equiv 1/r$ and $e \ll e_0$.
In this connection, it may be mentioned that there is some analogy with quantum electrodynamics where there are the observable electron mass
 $m_e$ and charge $e$,  as well as the corresponding bare mass $m_0 = \infty$ and charge $e_0 = \infty$.
 As in our case, in quantum electrodynamics, there are similar relationships $m_e \ll m_0$ and $e \ll e_0$, with the difference that in quantum electrodynamics
 $m_0$ and $e_0$ are infinite. However, it is so far unclear whether this analogy has a distinct physical meaning or this is just a coincidence.

\section{Discussion and conclusions}

In the present paper we have suggested two models of an electric charge, which for brevity was called an `electron'~\footnote{We have used here quotation marks, since it is so far unclear to what extent this model can be regarded as a model of a real electron. To clarify this question, additional theoretical studies are necessary, which would allow to get experimentally verifiable consequences.}. In both models, we have used the solutions obtained within $F(B^2)$ modified Weyl gravity of  Ref.~\cite{Dzhunushaliev:2020dom}. The models of Sections~\ref{m_n_0}
and Sec.~\ref{m_n_neq_0} ensures the electron charge, mass and spin equal to $1/2$ corresponding to a real electron.

The essence of the models is that one cuts out from Minkowski spacetime an interior region in the form of a torus, into which a spacetime described by a metric satisfying the equations of $F\left( B^2\right) $ modified Weyl gravity is embedded. To ensure the presence of a spin in such a {\it gravitating} system, it contains a Dirac spinor field. An interesting feature of  modified Weyl gravity under consideration is the presence of solutions to the Dirac equation possessing finite energy, electric charge, and spin $1/2$. We have demonstrated that these quantities can be chosen so that they would coincide with those for a real electron.

Another interesting feature of the `electron' is the fact that if one chooses the size of the torus cross-section $r\theta_0$ (on which the interior and exterior regions are joined) so that it would be much smaller than the radius of the axial circle of the torus $r$ (see Fig.~\ref{joining_torus}), for a distant observer such an object will look like a closed string with spinor degrees of freedom on it. This suggests that there might be some analogy to superstring theory, but such an analogy, apparently, need not imply a deeper significance,
since our stringlike object lives in a four-dimensional spacetime but not in a multidimensional one, and its Lagrangian cannot be reduced to the Nambu-Goto Lagrangian.

Notice also that, in the interior region, we employ the massless Dirac equation to describe the spinor field, but in the exterior region it is already necessary to use the Dirac equation with a mass term whose value is determined by the total mass of the spinor field contained inside the interior region.

The disadvantage of the model suggested here is the presence of discontinuities in the components of the metric tensor on the surface between interior and exterior regions of the `electron'. We note that this problem can, in principle, be solved rigorously only within the framework  of quantum gravity. It should be shown there that the interior region of such an object
can be approximately described by using modified Weyl gravity, and it is necessary to describe the region where the transition from quantum gravity to general relativity occurs. In a certain sense, this situation is similar to the presence of discontinuities in physical quantities at a shock wave whose structure cannot be already described by purely gasdynamic methods, and
one has to apply some simplified approximate models within the framework of kinetic theory.

Summarizing briefly the results obtained:
\begin{itemize}
\item Within modified Weyl gravity, the model of electric charge of a spin  $1/2$ is suggested.
\item The possibility of choosing the model parameters so as to get the mass and charge of a real electron is demonstrated.
\end{itemize}

\section*{Acknowledgments}
We gratefully acknowledge support provided by Grant No.~BR05236494
in Fundamental Research in Natural Sciences by the Ministry of Education and Science of the Republic of Kazakhstan. We are also grateful
to the Research Group Linkage Programme of the Alexander von Humboldt Foundation for the support of this research.

\appendix
\section{Spin operator in Hopf coordinates}
\label{spin_oper}

Let us calculate the projection of the operator of the total angular momentum  $n_i \hat M^i$ on the spinor, $n_i \hat M^i \psi$, where $n_i$ is a unit vector.
The operator  $\hat M_i$ is defined as
$$
	\hat M_i = \hat L_i + \hat S_i.
$$
Here, the operators of the orbital angular momentum $\hat L_i$ and the spin $\hat S_i$ in Cartesian coordinates are
$$
	\hat L_i = \epsilon_{i j k} x^j \hat p_k ,
\quad
	\hat S_i = \frac{1}{2}
	\begin{pmatrix}
		\sigma_i	&	0 \\
		0							&	\sigma_i
	\end{pmatrix},
$$
where $\sigma_i$ are the Pauli matrices and $\epsilon_{i j k}$ is the completely antisymmetric Levi-Civita symbol.

Consider the case of the projection of the operator $\hat M_i$ on the $z$-axis. Then the unit vector can be written in the form
$
	n_i = \left( 0, 0, 1 \right)
$. The vectors $\hat L_i, \hat S_i$, and $n_i$ in Hopf coordinates have the form
$$
	\hat{\tilde Q}_i = R^j_{\phantom{j} i} \hat Q_j , \quad
	\tilde n_i = R^j_{\phantom{j} i} n_j .
$$
Here $\hat{\tilde Q}_i = \hat{\tilde L}_i, \hat{\tilde S}_i $ and
$\hat{Q}_i = \hat{L}_i, \hat{S}_i $, and the tilde above the symbols denotes the corresponding operator
$\hat Q_i$ in Hopf coordinates. The transition matrix $R^i_{\phantom{i} j} = \partial x^i / \partial y^j$
between Cartesian and Hopf coordinates is
\begin{equation}
\begin{split}
	& R = 	\frac{r}{2
	\left[
		\cos \left(\frac{\theta }{2}\right)
		\sin \left(\frac{\varphi -\chi }{2}\right) - 1
		\right]^2
	}
\\
	&
	\begin{pmatrix}
	- \sin \left(\frac{\theta }{2}\right)
	\left(
		\cos \left(\frac{\theta }{2}\right) \cos (\varphi ) +
		\sin \left(\frac{\varphi +\chi }{2}\right)
	\right)
		&
	\cos \left(\frac{\varphi +\chi }{2}\right)
	\left[
		\cos \left(\frac{\theta }{2}\right) -
		\sin \left(\frac{\varphi -\chi }{2}\right)
	\right]
	&
	\sin \left(\frac{\theta }{2}\right)
	\left[
		\cos \left(\frac{\theta }{2}\right) \cos (\chi ) -
		\sin \left(\frac{\varphi +\chi }{2}	\right)
	\right]
	\\
	\sin \left(\frac{\theta }{2}\right)
	\left[
		\cos \left(\frac{\varphi +\chi }{2}\right) -
		\cos \left(\frac{\theta }{2}\right) \sin (\varphi )
	\right]
	&
	\sin \left(\frac{\varphi +\chi }{2}\right)
	\left[
		\cos \left(\frac{\theta }{2}\right) -
		\sin \left(\frac{\varphi -\chi }{2}\right)
	\right]
	&
	\sin \left(\frac{\theta }{2}\right)
	\left[
		\cos \left(\frac{\theta }{2}\right) \sin (\chi ) +
		\cos \left(\frac{\varphi +\chi }{2}\right)
	\right]
	\\
	- \cos \left(\frac{\theta }{2}\right)
	\left[
		\cos \left(\frac{\theta }{2}\right)-
		\sin \left(\frac{\varphi -\chi }{2}\right)
	\right]
	&
	- \sin \left(\frac{\theta }{2}\right)
	\cos \left(\frac{\varphi -\chi }{2}\right)
	&
	\cos \left(\frac{\theta }{2}\right)
	\left[
	\cos \left(\frac{\theta }{2}\right) -
	\sin \left(\frac{\varphi -\chi }{2}\right)
	\right]
	\end{pmatrix} ,
\label{5_50}
\end{split}
\nonumber
\end{equation}
and the unit vector in Hopf coordinates takes the form
\begin{equation}
\begin{split}
	& \tilde n_i = \frac{r}{
	2 \left[
		\cos \left(\frac{\theta }{2}\right) \sin \left(\frac{\varphi -\chi }{2}\right)
		- 1
	\right]^2
	}
	\\
	&
	\begin{pmatrix}
		- \cos \left(\frac{\theta }{2}\right)
		\left[
			\cos \left(\frac{\theta }{2}\right) -
			\sin \left(\frac{\varphi -\chi }{2}\right)
		\right] ,
	&
	- \sin \left(\frac{\theta }{2}\right)
	\cos \left(\frac{\varphi -\chi }{2}\right) ,
	&
	\cos \left(\frac{\theta }{2}\right)
	\left[
		\cos \left(\frac{\theta }{2}\right) -
		\sin \left(\frac{\varphi -\chi }{2}\right)
	\right]
	\end{pmatrix} .
\label{5_60}
\end{split}
\nonumber
\end{equation}
This enables us to calculate the projections of the operators
$\hat{\tilde L}_i$ and $\hat{\tilde S}_i $ on the $z$-axis in the form
$$
	n^i \hat{\tilde L}_i = \hat{\tilde L}_z =
	- i \left(
		\partial_\chi + \partial_\varphi
	\right) ,
\quad
	n^i \hat{\tilde S}_i = \hat{\tilde S}_z =
	\frac{1}{2}
	\begin{pmatrix}
			\sigma_z	&	0 \\
			0							&	\sigma_z
	\end{pmatrix} .
$$


\begin{thebibliography}{99}

\bibitem{ablo}
M. Abraham,
Phys. Z. \textbf{4}, 57 (1902);
Ann. der Phys. \textbf{10}, 105 (1903).

\bibitem{ablo1}
H. A. Lorentz,
in ``\textit{Encyclopadie der Mathematischen Wissenschaften}'', Vol. 5,
Leipzig: Teubner (1905-1922), p 145.

\bibitem{wheelbook}
J. A. Wheeler,
{\it Geometrodynamics} (New York: Academic Press, 1962).

\bibitem{Born:1934gh}
  M.~Born and L.~Infeld,
  Proc.\ Roy.\ Soc.\ Lond.\ A {\bf 144}, no. 852, 425 (1934).

\bibitem{heis}
	W.~Heisenberg, {\it Introduction to the Unified Field Theory of Elementary Particles} (Interscience Publishers, London, 1966).

\bibitem{Markov:1970hv}
  M.~A.~Markov,
  Annals Phys.\  {\bf 59}, 109 (1970).

\bibitem{Markov:1972sc}
  M.~A.~Markov and V.~P.~Frolov,
  Teor.\ Mat.\ Fiz.\  {\bf 13}, 41 (1972).

\bibitem{Bronnikov:1979ex}
  K.~A.~Bronnikov, V.~N.~Melnikov, G.~N.~Shikin, and K.~P.~Staniukowicz,
  Annals Phys.\  {\bf 118}, 84 (1979).

\bibitem{Datzeff:1980tg}
  A.~B.~Datzeff,
  Phys.\ Lett.\  A {\bf 80}, 6 (1980).

\bibitem{Bonnor:1989iw}
  W.~B.~Bonnor and F.~I.~Cooperstock,
  Phys.\ Lett.\  A {\bf 139}, 442 (1989).

\bibitem{Davidson:1992kr}
  A.~Davidson and U.~Paz,
  Phys.\ Lett.\ B {\bf 300}, 234 (1993).

\bibitem{Davidson:1990ez}
A.~Davidson and E.~Guendelman,
Phys. Lett. B \textbf{251} (1990), 250-253.

\bibitem{Herr:1994}
L. Herrera and V. Varela,
Phys. Lett. A \textbf{189}, 11 (1994).

\bibitem{Zaslavskii:2010yi}
E. Ayon-Beato and A. Garcia,
Phys. Rev. Lett. \textbf{80}, 5056 (1998).

\bibitem{Finst:1999}
  F.~Finster, J.~Smoller, and S.~-T.~Yau,
  Phys.\ Lett.\ A {\bf 259}, 431 (1999).

\bibitem{Bronn:2000}
K.~A.~Bronnikov,
  Phys.\ Rev.\ Lett.\  {\bf 85}, 4641 (2000).

\bibitem{Habib:2009}
S.~Habib Mazharimousavi and M.~Halilsoy,
  Phys.\ Lett.\  B {\bf 678}, 407 (2009).

\bibitem{Zaslav:2010}
  O.~B.~Zaslavskii,
  Grav.\ Cosmol.\  {\bf 16}, 168 (2010).

\bibitem{Dzhunushaliev:2016bnh}
  V.~Dzhunushaliev, A.~Makhmudov, and K.~G.~Zloshchastiev,
  Phys.\ Rev.\ D {\bf 94}, no. 9, 096012 (2016).

\bibitem{Finster:1998ws}
  F.~Finster, J.~Smoller, and S.~-T.~Yau,
  Phys.\ Rev.\  D {\bf 59}, 104020 (1999).

\bibitem{KNold}
B. Carter,
Phys. Rev. \textbf{174}, 1559 (1968).

\bibitem{KNold1}
W. Israel,
Phys. Rev. D \textbf{2}, 641 (1970).

\bibitem{KNold2}
G.C. Debney, R.P. Kerr, and A.Schild,
J. Math. Phys. \textbf{10}, 1842 (1969).

\bibitem{KNold3}
A. Burinskii,
Sov. Phys. JETP \textbf{39}, 193 (1974);
Russian Phys. J. \textbf{17}, 1068 (1974);
Phys.\ Rev.\  D {\bf 67}, 124024 (2003);
Grav. Cosmol. \textbf{14}, 109 (2008).

\bibitem{KNold4}
D. Ivanenko and A.Ya. Burinskii,
Russian Phys. J. \textbf{18}, 721 (1975).

\bibitem{KNold5}
C. A. Lopez,
Phys. Rev. D \textbf{30}, 313 (1984);
Gen. Rel. Grav. {\bf 24}, 285 (1992).

\bibitem{KNold6}
M. Israelit and N. Rosen, Gen. Rel. Grav. {\bf 27}, 153 (1995).
\bibitem{KNold7}
  H.~I.~Arcos and J.~G.~Pereira,
  Gen.\ Rel.\ Grav.\  {\bf 36}, 2441 (2004).

\bibitem{Burinskii:2011id}
  A.~Burinskii,
  ``Gravity versus Quantum theory: Is electron really pointlike?,''
	arXiv:1104.0573 [hep-ph].

\bibitem{Dzhunushaliev:2020dom}
V.~Dzhunushaliev and V.~Folomeev,
``Spinor field solutions in $F\left(B^2\right)$ modified Weyl gravity,''
[arXiv:2003.02646 [gr-qc]].

\bibitem{Nojiri:2010wj}
  S.~Nojiri and S.~D.~Odintsov,
  Phys.\ Rept.\  {\bf 505}, 59 (2011).

\bibitem{Nojiri:2017ncd}
  S.~Nojiri, S.~D.~Odintsov, and V.~K.~Oikonomou,
  Phys.\ Rep.\  {\bf 692}, 1 (2017).

\bibitem{Lawrie2002}
I.~Lawrie, {\it A unified grand tour of theoretical physics} (Institute of Physics
Publishing, Bristol and Philadelphia, 2002).

\end{thebibliography}
\end{document}